\documentstyle[prl,aps,twocolumn]{revtex}
\input{epsf}

\def\tp{${\cal T}^{(P)}$}
\def\tsq{${\cal T}^{*}$}
\def\ts2f{${\cal T}^{*(2F)}$}
\def\tsa4{${\cal T}^{*(A_4)}$}
\def\tp1r{${\cal T}^{*((P1)r)}$}
\def\tsp1{${\cal T}^{*(P1)}$}

\def\cktsa4{${\cal C}^K_{{\cal T}^{*(A_4)}}$}
\def\cstsa4{${\cal C}^s_{{\cal T}^{*(A_4)}}$}

\def\es{${{\mbox{I}}\!{\mbox{{\bf E}$_\perp $} }}$}
\def\qd6{$q_{D_6}$}
\def\d6{$D_6$}

\def\z6{$Z\!\!\!Z^6$}

\def\iAPM{\mbox{$i$-Al-Pd-Mn}}

\begin{document}

\draft
\title{Tiling of the five-fold surface of
Al$_{70}$Pd$_{21}$Mn$_{9}$}
\author{J. Ledieu and
R. McGrath\footnote{Author for correspondence.
e-mail:mcgrath@liv.ac.uk;
Phone:+44 151 794 3873;
Fax:+44 151 708 0662}}
\address{Surface Science Research Centre,
\\The University of
Liverpool, Liverpool L69 3BX, UK}
\author{R.D. Diehl}
\address{Department of Physics, Pennsylvania State University,
University Park, PA 16802, USA}
\author{T.A. Lograsso and D.W. Delaney}
\address{Ames Laboratory, Iowa State University,
Ames, IA 50011, USA}
\author{Z. Papadopolos and G. Kasner}
\address{Institut f\"{u}r Theoretische Physik,
Universit\"{a}t Magdeburg, PSF 4120, D-39016 Magdeburg, Germany}
\date{April 05, 2001}
\maketitle

\begin{abstract}

The nature of the five-fold surface of 
Al$_{70}$Pd$_{21}$Mn$_{9}$ has been investigated using scanning 
tunnelling microscopy.  From high resolution images of the 
terraces, a tiling of the surface has been constructed using 
pentagonal prototiles.  This tiling matches the bulk
model of Boudard \textit{et al.} (J.  Phys: Cond.  Matter
\textbf{4}, 10149 (1992)), which allows us to elucidate the 
atomic nature of the surface.  Furthermore, it is consistent 
with a Penrose tiling \tp1r \ obtained from the geometric model 
based on the three-dimensional tiling \ts2f.  The results 
provide direct confirmation that the five-fold surface 
of \iAPM \ is a termination of the bulk structure.

\end{abstract}


Since their discovery \cite{Shechtman84}, quasicrystals have 
extended the boundaries of our knowledge, most strikingly in
the redefinition of the crystal undertaken by the International 
Union of Crystallography in 1991 \cite{IUCR92}.  The reason 
for this lies in their unusual aperiodic structure, which in 
the case of \iAPM \ and \mbox{$i$-Al-Cu-Fe} has been 
described mathematically with reference to a six-dimensional 
%
%
lattice \d6 \ \cite{kg,e}.
The fact that a three-dimensional atomic model \cite{kg,e} 
can be based on a three-dimensional tiling projected from 
the \d6 \ lattice~\cite{KPL} leads us to expect the five-fold 
planes of the model to be related to a two-dimensional 
Penrose-like tiling \cite{Penrose74,GS}.

The unusual tribological behavior observed for quasicrystals 
raises questions concerning the nature of their 
surfaces \cite{Dubois93}. Systematic studies by Gellman and 
coworkers indicate that the static friction coefficient for 
\iAPM \ (on itself) is lower than that of most pure metals, 
and the slip-stick behavior commonly observed on
crystalline surfaces is not present \cite{Ko99}.  A complete
understanding of these observations requires a knowledge of the
quasicrystal surface structure \cite{Persson99}.  It can not be
assumed \textit{a priori} that a quasicrystal surface is 
aperiodic itself or that it reflects a perfect truncation of 
the bulk structure. If this is the case, however, we would 
expect the structure of that surface to reflect the symmetry 
of a two-dimensional Penrose tiling \cite{KPL,Penrose74,GS}.  
Until now, however, this direct link between theory and 
experiment has not been made.

This is partly because the aperiodic nature of quasicrystals 
makes it difficult to determine their surface structure.  
Surface diffraction techniques can not be exploited to achieve 
a full structural determination as they rely on a formalism 
developed largely for periodic structures 
\cite{Gierer97,Gierer98}.  Scanning probe microscopies offer 
an alternative, but even with these methods atomic resolution 
has so far proved elusive.  It has been variously
suggested that this is an inherent limitation of the electronic
structure of these surfaces \cite{Shen99} or a consequence of
defect-like protrusions observed in all studies to date
\cite{Shen99,Schaub94a,Ledieu99a,Ledieu99b}.  In previous
work, we introduced an approach based on tiling of scanning 
tunnelling microscopy (STM) images using regions of high 
contrast as vertices~\cite{Ledieu99a}.  Though this approach 
produced partial tilings, the presence of large protrusion 
defects on the surface introduced breaks in the tiling and 
comparison with models of the surface was not possible.

 In this study we carry this tiling approach to fruition, 
using a combination of experimental and theoretical methods.  
A refined surface preparation technique has led to terraces 
which are free from the protrusions found in previous STM
studies~\cite{Shen99,Schaub94a,Ledieu99a,Ledieu99b}.  This in
turn has led to better resolution STM images which together 
with the structural perfection, allows us to derive an 
experimental tiling over a lateral range of $\sim$100 \AA{}.  
By comparison with the experimentally derived bulk model 
of Boudard \textit{et al.} \cite{Boudard92}, we identify 
possible structural entities on the surface.  We also 
demonstrate that the experimental tiling matches the 
two-dimensional tiling \tp1r, derived from a well
established geometric model of the bulk \cite{kg,e}.

The quasicrystal sample was grown using the Bridgman method 
and polished using 6$\mu$m, 1$\mu$m and 0.25$\mu$m diamond 
paste on Texmet cloth for one hour.  The ultra-high vacuum 
(UHV) preparation consisted of five sputter/anneal cycles.  
The sputtering angle was at grazing incidence 
(20$^{\circ}$-30$^{\circ}$); Ar gas was used at 500~eV and 
each sputter lasted 90 minutes.  Each annealing was to 970~K
for 120-150 minutes.  After this procedure, the LEED pattern 
had five-fold rotational axes and exhibited a low background, 
with sharp peaks.  We note that this temperature is at the 
upper end of the range known to produce quasicrystalline 
surfaces \cite{Thiel99}.

%
%
\begin{figure}[ht]
\begin{center}
     \epsfxsize=70mm
     \epsffile{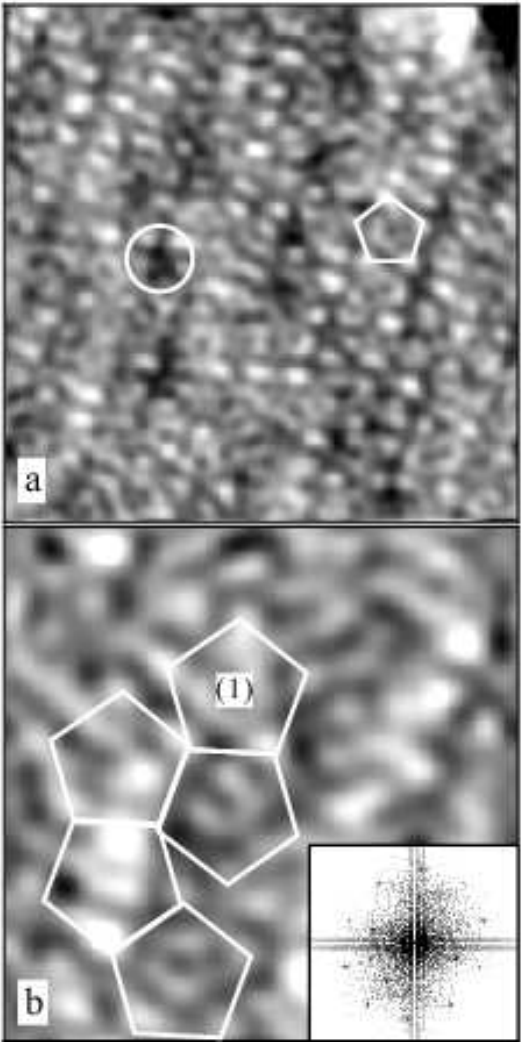}
      \caption{(a) 100 $\times$ 100 \AA{}$^{2}$ high resolution 
                   STM image of a flat terrace (V = 1 V, 
                   I = 0.3 nA).  A pentagonal hole (circled) and 
                   a pentagon have been outlined.
               (b) 50 $\times$ 50 \AA{}$^{2}$ high 
                   resolution STM image (V = 1 V, I = 0.3 nA).  
                   Several pentagons have been outlined.  A Fast 
                   Fourier Transform (FFT) of the image in (b) 
                   is shown as an inset.}
     \label{fig:flatterrace}
\end{center}
\end{figure}

This preparation procedure leads to very large flat terraces 
(up to \mbox{1500 \AA{}} wide).  Images from the terraces were 
then obtained which display a higher resolution than those in 
any previously published STM
work~\cite{Shen99,Schaub94a,Ledieu99a,Ledieu99b}.
Fig.~\ref{fig:flatterrace}(a) shows a 
100 $\times$ 100 \AA{}$^{2}$ image of a flat terrace 
with features of atomic size (\mbox{2-3 \AA{}}).  The 
corrugation across the terraces (see
Fig.~\ref{fig:flatterrace}(a)) is \mbox{$<$ 1 \AA{}}.  
Large areas of the terraces are found without any protrusions, 
which allows the tip to scan the surface more closely leading 
to improved resolution. Pentagonal areas of dark contrast 
(`holes') are observed as in previous 
work \cite{Shen99,Schaub94a,Ledieu99a,Ledieu99b} but
now with much improved definition - see the circled area 
in Fig.1(a). A Fast Fourier Transform (FFT) 
(inset on Fig.  1(b)) shows apparent ten-fold symmetry, 
consistent with a five-fold quasicrystalline surface.  
Autocorrelation patterns from such images (not shown) reveal
ten-fold symmetry and a high degree of order.

A notable feature of the surface is that areas of bright 
contrast may be joined by lines to form regular pentagons, 
with edges of length 8.0$\pm0.3$ \AA{}.  One such pentagon 
is outlined in Fig.~1(a).  Using these pentagons as basic 
building blocks 
%
%
we then construct a tiling 
of the surface.  This procedure is illustrated in Fig.~1(b) 
on a 50 $\times$ 50 \AA{}$^{2}$ image of the surface.  
Starting with the pentagon labelled (1), we look for pentagons 
sharing edges which also have areas of bright contrast at 
their vertices.  In this way, a tiling of the surface can be 
constructed as shown in Fig.~2.  Within this patch, the acute 
rhombus, crown and pentagonal star tiles then appear naturally 
in addition to the pentagonal tiles.  This tiling has
the appearance of the Penrose tiling (P1) \cite{GS}.  
We estimate that 93\% of the 119 vertices of the tiling 
patch coincide with areas of higher contrast; 
of the remaining 7\% which coincide with darker areas
the majority are on the perimeter of the tiling where any 
built-in strain due to slight mismatch will be maximum.  
Another possibility is the presence of vacancy defects 
(common on metal surfaces), although a slight distortion 
of the image toward the top edge due to piezo drift
cannot be excluded.

%
%
\begin{figure}[ht]
\begin{center}
      \epsfxsize=70mm
      \epsffile{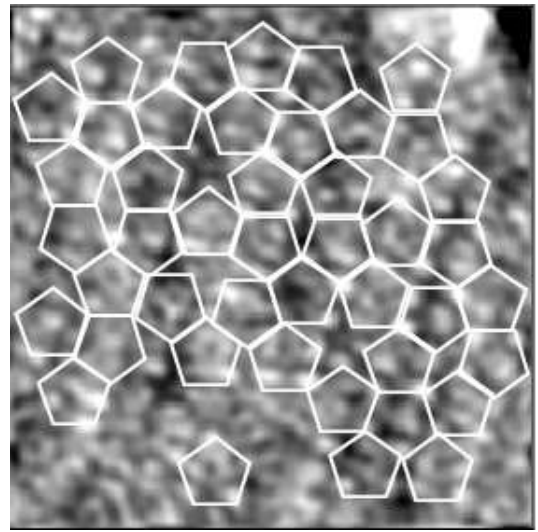}
      \caption{Tiling of Fig.~\ref{fig:flatterrace}(a) 
               derived as described in the text.}
     \label{fig:tilinghighres}
\end{center}
\end{figure}

\begin{figure}[ht]
\begin{center}
      \epsfxsize=70mm
      \epsffile{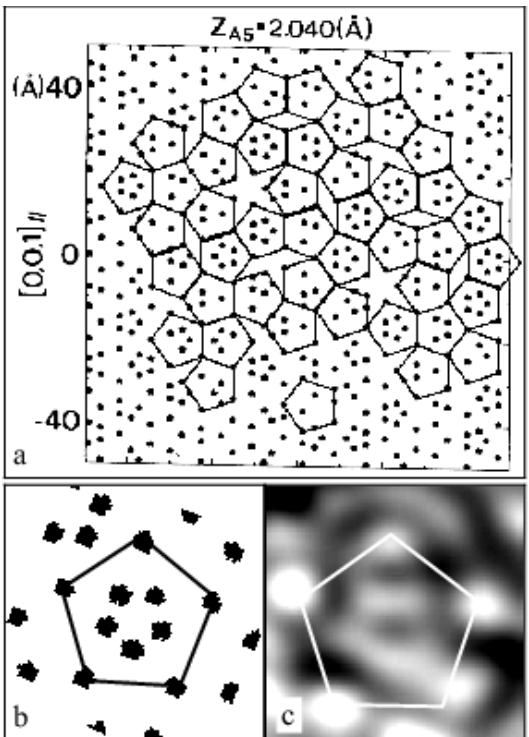}
      \caption{(a) Experimentally constructed tiling of
                   Fig.~\ref{fig:tilinghighres} superimposed 
                   on a plane of the model of 
                   Boudard \textit{et al.} 
               (b) corresponds to a 
                   20 $\times$ 20 \AA{}$^{2}$ region of 
                   this plane. 
               (c) represents a 20 $\times$ 20 \AA{}$^{2}$
                   region of Fig.~\ref{fig:flatterrace}(a).}
     \label{boudard}
\end{center}
\end{figure}

This tiling is not necessarily unique, and the tiles themselves
contain internal structure; but this is precisely what is to be
expected for this surface, as can be shown by comparison with 
the experimentally derived model of Boudard \textit{et al.}, 
based on x-ray and neutron scattering 
measurements \cite{Boudard92}. Fig.~\ref{boudard}(a) shows one 
plane from this model, which contains only aluminum atoms.  
The experimentally derived tiling of Fig.~2 is shown 
superimposed on this plane without scaling.
Fig.~\ref{boudard}(b) shows that within the pentagons 
described above, other atoms are also expected to be present, 
which in turn will give rise to high contrast areas within 
the pentagons on the STM images. This is demonstrated in 
Fig.~\ref{boudard}(c) \cite{Boudard92}.

%
%

The close correspondence of the experimentally derived tiling 
with the surface termination of the Boudard model allows us to 
speculate on the structure of the atomic units on the surface.  
The plane from the Boudard model consists entirely of
Al atoms, suggesting that the surface measured is Al-rich.  
This is in agreement with the results of 
LEED \cite{Gierer97,Gierer98}.  Recent LEIS
measurements of the surface \cite{Bastasz01} have
identified a characteristic distance of 7.5$\pm$0.1 \AA{}; 
our results suggest that this corresponds to the Al-Al 
distance along the edges of the pentagons.

We now show that this tiling is also contained within the 
Katz-Gratias-Elser geometric model~\cite{kg,e}.  
The geometric model consists of the three-dimensional 
quasicrystalline tiling \ts2f \ derived from a six-dimensional 
lattice, 
face-centered hypercubic lattice $D_{6}$.  This 
tiling is decorated by Bergman (and automatically Mackay) 
polytopes to give the atomic positions~\cite{KPL}.   All of the 
vertices of the \ts2f\  tiling can be embedded in a sequence 
of planes orthogonal to a five-fold symmetry axis of an 
icosahedron~\cite{Kasner99}.

\begin{figure}[ht]
\begin{center}
\epsfxsize=86mm
\epsffile{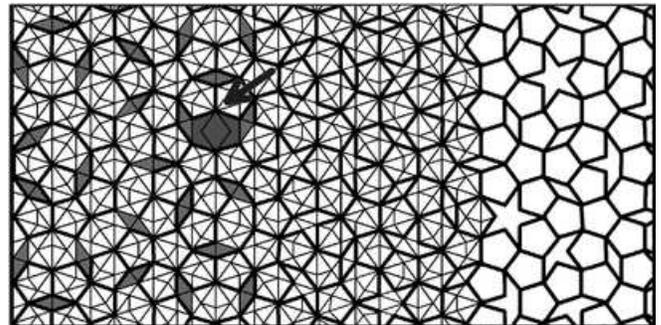}
\caption{Left side: The tiling \tsq \ of the plane by the acute
                    rhombus, pentagon and hexagon, locally 
                    derived from \tsa4. 
            Centre: The construction  of the tiling \tp1r.
        Right side: The tiling \tp1r\ without the content of 
                    the golden triangles.}
\label{taustar}
\end{center}
\end{figure}

In some of these planes a quasiperiodic tiling \tsa4\ 
appears (~\cite{Kasner99} and refs. therein).  The prototiles 
in \tsa4 \ are golden triangles.  The edges of the triangles 
in the tiling are parallel to the two-fold symmetry axes of 
an icosahedron (``two-fold directions") and are of two lengths 
related by the golden ratio $\tau$.  As an intermediate step, 
we locally derive the tiling \tsq \ with pentagon, acute 
rhombus and hexagon as prototiles from the quasilattice of the 
tiling \tsa4, as shown in Fig.~\ref{taustar}(left side).  
The tiling has an inflation factor $\tau$.

%
%

Keeping all acute rhombuses from the tiling \tsq, we replace 
each hexagon by two overlapping pentagons.  Now we 
{\em randomly} choose one of the pentagons from each 
overlapping pair, and unify the rest of the hexagon(s) with 
the neighboring acute rhombus.  This process is indicated by 
an arrow in the centre section of Fig.~\ref{taustar}.  In
this way we obtain either a crown or a pentagonal star to 
replace the rhombus, and the result is a tiling \tp1r, 
see the right-hand side of Fig.~\ref{taustar}.  
The tiling \tp1r\ is a partly random variant of the 
Penrose (P1) tiling, \tsp1~\cite{GS,Ni89}.

The tiling \tp1r \ clearly matches the geometry of the 
experimentally derived tiling shown in 
Fig.~\ref{fig:tilinghighres}.  We now compare the edge-length 
of the tiles to the experimental value of 8.0$\pm$0.3\AA{}.  
The scaling for the tilings derived above from the model is 
implicit from previous work ~\cite{Kasner99,Papadopolos99}: 
the edges of the tilings \tsa4, \tsq, and \tp1r\ are of length 
12.553~\AA{}.  However by an investigation of the window in 
perpendicular space \es \ of the atomic positions for the
plane in the model which we estimate most closely corresponds 
to the experimental one we may show that the minimal edge 
length for a P1($r$) tiling  
%
%
is $\tau^{-1}$12.553 \AA{}=7.758 \AA{}. Therefore within 
the geometric model, in a plane that corresponds to that 
of Fig.~2, a tiling such as in Fig.~\ref{taustar} exists 
with an edge length 7.758 \AA{}, in excellent agreement 
with the experimental observations.

Our data and analysis provide dramatic confirmation of the  
bulk termination of this surface which has been suggested 
by other workers. The LEED analysis of Gierer and co-workers 
indicated that the surface is consistent with the bulk 
quasicrystallinity and is an Al 
termination~\cite{Gierer97,Gierer98}.  X-ray photoelectron 
diffraction (XPD) studies are also consistent with a 
quasicrystalline surface nature~\cite{Naumovic99}.  Previous 
scanning tunnelling microscopy (STM) studies have all 
presented similar images of the quasicrystalline 
surface ~\cite{Shen99,Schaub94a} having a lower resolution 
than those presented here (the degree of resolution can be put
on a quantitative basis using radial distribution functions 
(RDF) from autocorrelation patterns (not shown)).  
Shen~\textit{et al.}, using an autocorrelation analysis 
showed that the surface structure is consistent with a bulk 
structure based on truncated pseudo-Mackay icosahedra or 
Bergman clusters~\cite{Shen99}.  
Schaub~\textit{etal.}~\cite{Schaub94a} interpreted their 
STM images in terms of an Ammann pentagrid model with 
Fibonacci relationships between structural elements within 
the terraces and across steps on the surface.  Later, these 
measurements were shown to be in correspondence
with the Katz-Gratias-Elser geometric model~\cite{kg,e} 
for the atomic positions~\cite{Kasner99,Papadopolos99}.  
However the lower resolution of these measurements precluded 
their analysis using the tiling approach described in this paper.

The ability to prepare surfaces having the structural perfection
observed here will facilitate the precise characterization of 
ordered molecular adsorption,  friction  and adhesion on 
quasicrystal surfaces. In previous work on the adsorption of 
C$_{60}$ molecules of the \iAPM \ surface, although local areas
were found in which the molecules had Fibonacci scaling 
relationships, on a larger scale the correlation was broken 
by the presence of defects~\cite{Ledieu01}. Surfaces of the 
quality described above should enable the formation of better 
ordered overlayers.

To summarize, very high resolution STM images of flat
terraces of the five-fold Al$_{70}$Pd$_{21}$Mn$_{9}$
surface have been presented.  A tiling of the surface
based on pentagons of edge \mbox{8.0 $\pm$ 0.3\AA{}}
has been experimentally derived.  This tiling is shown to be
consistent with the geometric model based on the \ts2f \ tiling, 
and with the experimentally derived model of 
Boudard \textit{et al.} \cite{Boudard92}.
These results point clearly to a bulk termination of this
surface and lend support to the bulk models of this complex
material.

The EPSRC (Grant numbers GR/N18680 and GR/N25718), NFS (Grant
number DMR-9819977) and DFG (Grant number KA 1001/4-2)
are acknowledged for funding.

\end{document}